\documentclass[iop]{emulateapj}

\usepackage{txfonts}
\usepackage{graphicx}
\usepackage{graphicx}
\usepackage{soul}

\usepackage{natbib}
\newcommand{\ks}{\mbox{km~s$^{-1}$}}

\usepackage{natbib}

\begin{document}
\title{Observations and simulations of the N\lowercase{a} \lowercase{{\sc{i}}} D$_{1}$ line profiles in an M-class solar flare}
\vskip1.0truecm
\author{D.~Kuridze$^{1,5}$, M.~Mathioudakis$^{1}$, D. J. Christian$^2$, A.~F.~Kowalski$^4$, D. B. Jess$^1$, S. D. T. Grant$^1$, T. Kawate$^1$,
P.~J.~A.~Sim\~{o}es$^2$, J.~C.~Allred$^4$, \& F. P. Keenan$^1$}
\affil
{$^1$Astrophysics Research Centre, School of Mathematics and Physics, Queen's University Belfast, BT7 1NN, Northern Ireland, UK.}
\affil{$^2$Department of Physics and Astronomy, California State University, Northridge, CA 91330, USA.}
\affil{$^3$SUPA School of Physics and Astronomy, University of Glasgow, Glasgow G12 8QQ, U. K.}
%\affil{$^3$Institute of Theoretical Astrophysics, University of Oslo, P.O. Box 1029 Blindern, N-0315 Oslo, Norway.}
\affil{$^4$ NASA/Goddard Space Flight Center, Code 671, Greenbelt, MD 20771.} 
\affil{$^5$Abastumani Astrophysical Observatory at Ilia State University, 3/5 Cholokashvili avenue, 0162 Tbilisi, Georgia}
\date{received / accepted }

%%%%%%%%%%%%%%%%%%%%%%%%%%%%%%%%%%%%%%%%%%%%%%%%%%%%%%%%%%%%%%%%%%%%%%%%%

\begin{abstract}

We study the temporal evolution of the Na {\sc{i}} D$_{1}$ line profiles in the M3.9 flare SOL2014-06-11T21:03~UT, 
using high spectral resolution observations 
obtained with the IBIS instrument on the Dunn Solar Telescope combined with radiative hydrodynamic simulations. Our results show a significant increase in line core and wing intensities during the flare. 
The analysis of the line profiles from the flare ribbons reveal that the Na {\sc{i}} D$_{1}$ line has a central reversal with 
excess emission in the blue wing (blue asymmetry). We combine RADYN and RH simulations to synthesise 
Na {\sc{i}} D$_{1}$ line profiles of the flaring atmosphere and find good agreement with the observations. 
Heating with a beam of electrons modifies the radiation field in the flaring atmosphere and excites 
electrons from the ground state $\mathrm{3s~^2S}$ to the first excited state $\mathrm{3p~^2P}$, 
which in turn modifies relative population of the two states.
The change in temperature and the population density of the energy states make the sodium line profile revert from absorption into emission. 
Analysis of the simulated spectra also reveals that the Na {\sc{i}} D$_{1}$ flare profile asymmetries are produced by the velocity gradients generated %and opacity effects 
in the lower solar atmosphere. 

\end{abstract}

%%%%%%%%%%%%%%%%%%%%%%%%%%%%%%%%%%%%%%%%%%%%%%%%%%%%%%%%%%%%%%%%%%%%%%%%%

\section{Introduction}

The lower solar atmosphere is key to our understanding of solar flares, 
as the vast majority of the flare radiative energy originates in the chromosphere and photosphere. 
Chromospheric radiation is dominated by the optically thick lines of hydrogen, calcium and magnesium which provide diagnostics on flare dynamics.  
One of the main characteristics of the flaring chromosphere is the centrally-reversed H$\alpha$ emission profile with asymmetric red and blue wings. The asymmetries are attributed to 
the downflow and upflow of plasma, triggered by the deposition of non-thermal energy, 
which are also known as chromospheric condensation and evaporation \citep{1984SoPh...93..105I,1989ApJ...341.1088W,1992ApJ...401..761D}. These processes can be very effective tracers for the velocity field of the flaring atmosphere. 
\cite{1999ApJ...521..906A} computed flare time-dependent H$\alpha$ and Ca {\sc{ii}} K line profiles with the radiative-hydrodynamic code 
\citep[RADYN;][]{1997ApJ...481..500C}, and showed that the asymmetries could be produced by the strong velocity gradients generated during the flare. 
These gradients create differences in the opacity between the red and blue wings of H$\alpha$ and their sign 
determines whether the asymmetric emission appears to the blue or red side of the line profile.   
Recently, \cite{2015ApJ...813..125K} made a direct comparison of the observation and RADYN simulation for the evolution of the H$\alpha$ profile. 
They showed that the steep velocity gradients in the flaring chromosphere modify the wavelength of the central reversal in H$\alpha$. 
The shift in the wavelength of maximum opacity to shorter and longer wavelengths generates the red and blue asymmetries, respectively.

%Observations of solar flares in the deeper layers of the atmosphere is still very important and challenging part of the flare studies. 
%With modern state-of-the-art instrumentation increased evidences of the photospheric signatures during the flares have been detecting in the absorption line such as Fe %{\sc{i}}, the molecular G-band and white-light blue continuum lines (4170 \& 3501 $\AA$) 
%as well as white-light optical continuum \citep{2015SoPh..290.3487K}.
The exact formation height of the Na {\sc{i}} D$_1$ line core is difficult to estimate from  
observations, and with simulations indicating it is formed below the formation height of the H$\alpha$, Ca {\sc{ii}} H \& K, and the Ca {\sc{ii}} infrared (IR) triplet line cores.  
\cite{2008AN....329..494S} explored the formation height of the Na {\sc{i}} D lines by calculating the response functions of their 
profiles to p-mode power variations. They concluded that the line has its origin in a wider region, from the photosphere up to the lower chromosphere at height of around 800 km.

Observations and simulations performed for the quiet solar atmosphere have revealed
that the Na {\sc{i}} D$_1$ core brightness samples the magnetic bright points in the solar photosphere and is strongly affected by 3D resonance scattering 
\citep{2002ESASP.477..147M,2010ApJ...709.1362L,2010ApJ...719L.134J}. These studies suggest that the line core emission in the quiet Sun originates in the upper photosphere and lower chromosphere.   

\begin{figure*}[t]
\begin{center}
\includegraphics[width=17.4cm]{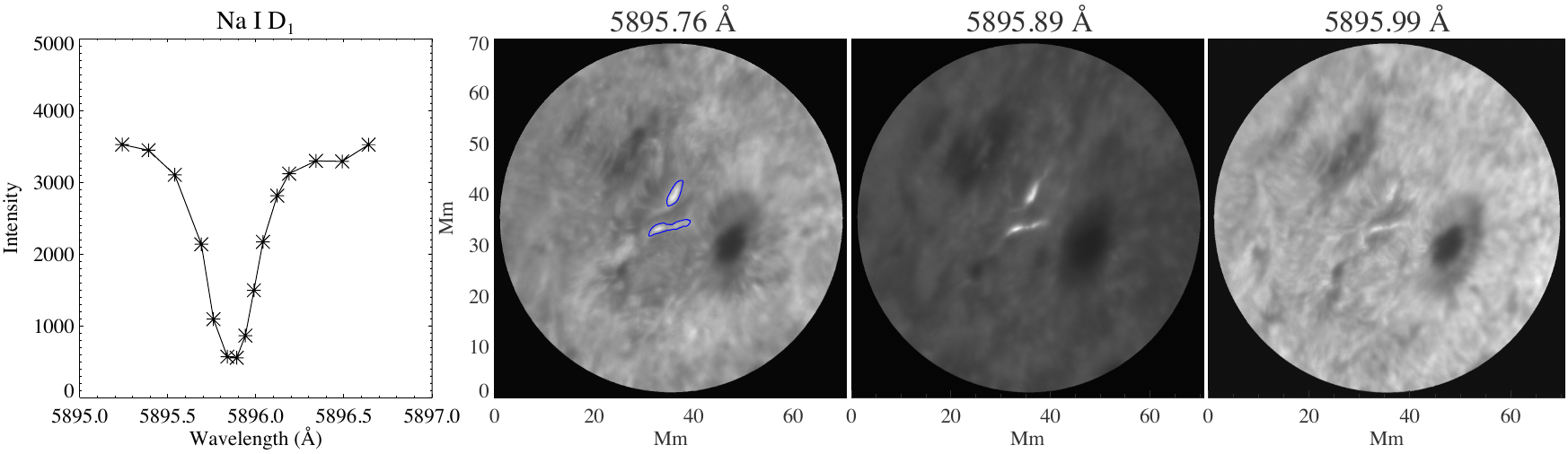}
\end{center}
\caption{(Left) The IBIS Na {\sc{i}} D$_1$ line profile of a quiet solar region. Asterisks show the spectral positions selected for the IBIS line scan. 
The full black line shows the mean spectrum averaged over the quiet Sun FOV. (Right) Na {\sc{i}} D$_1$ wing and core images at selected wavelength positions. 
The blue contours show the upper and lower flare ribbons analysed in this paper. Contours indicate the 50\% level of the intensity maximum.}
\label{fig1}
\end{figure*}

There is a lack of flare observations and modelling in the Na {\sc{i}} D lines, with the exception of some early work \citep{1967SoPh....1..389G,1990A&AS...84..601F,1996A&A...310L..29M}. 
The most recent observations of solar flares in Na {\sc{i}} D$_1$ have been reported by \cite{2010SoPh..263..153C}, 
who analysed the D$_1$ and D$_2$ line intensities observed with the GOLF spectrophotometer onboard SOHO. They found that the intensities of these lines integrated over the solar disk are increased during flares. 
However, the GOLF instrument can only record intensities in the single wavelength positions of $\pm$0.108~{\AA} from the line core
and does not allow for full spectral line profiles.

In this paper we present high temporal, spatial and spectral resolution observations of an M3.9 solar flare in Na {\sc{i}} D$_1$.  
Multi-wavelength observations of this M3.9 flare are also presented in \cite{christian2016}.
We study the evolution of the line profiles of the flare ribbons, and compare our findings with synthesised profiles obtained with a radiative hydrodynamic simulation.   
Motivated by the close match between simulations and observations, we investigate the formation of centrally-reversed, asymmetric Na {\sc{i}} D$_1$ line profiles using synthetic spectra. 
The line contribution functions and the velocity field in the simulated atmosphere allow 
us to investigate the nature of the observed line asymmetries. 

%%%%%%%%%%%%%%%%%%%%%%%%%%%%%%%%%%%%%%%%%%%%%%%%%%%%%%%%%%%%%%%%%%%%%%%%%

\section{Observations and data reduction}
\label{sect:setup}

The observations presented in this work were obtained between 19:20 and 21:27 UT on 2014 June 11
with the Interferometric Bidimensional Spectrometer \citep[IBIS;][]{2006SoPh..236..415C}, mounted at the Dunn Solar Telescope (DST) at the National Solar Observatory, NM, USA.
%to correct wavefront deformations in real time. 
High-order adaptive optics were applied throughout the observations \citep{2004SPIE.5490...34R}.
IBIS acquired Na {\sc{i}} D$_1$ spectral imaging with a spatial sampling of 0.0976 $''$ pixel$^{-1}$. The Na {\sc{i}} D$_1$ line scan consists of 15 positions 
%{\bf(?0.906, ?0.543, ?0.362, 0.000, 0.362, 0.543, +0.906  {\AA} from line core)}
corresponding to a velocity range of -37 to +34 \ks  
%sampled with 0.2 ${\AA}$ steps from line core 
(left panel of Figure~\ref{fig1}). 
Speckle reconstruction was applied to the data \citep{2008A&A...488..375W}, 
utilizing 10 $\rightarrow$ 1 restorations.  %A post reconstruction full spectral scan had an acquisition time of a 22.709 s, which is the effective cadence of the timeseries.
A full post reconstructed IBIS scan through the Na {\sc{i}} D$_1$ absorption profile had an acquisition time of 22.709 s, which is the effective cadence of the time-series and
includes a blueshift correction required due to the use of classical etalon mountings \citep{2008A&A...480..515C}.

The flare was also observed by RHESSI \citep{2002SoPh..210....3L}. 
Hard X-ray (HXR) spectral analysis was performed using OSPEX \citep{2002SoPh..210..165S}
to estimate the power $P_\mathrm{nth}$ 
deposited in the chromosphere by the nonthermal electrons, assuming the thick-target model \citep{1971SoPh...18..489B}.

%%%%%%%%%%%%%%%%%%%%%%%%%%%%%%%%%%%%%%%%%%%%%%%%%%%%%%%%%%%%%%%%%%%%%%%%%

\section{Analysis and results}

\begin{figure}[t]
%\vspace{3mm}
\begin{center}
\includegraphics[width=8.0 cm]{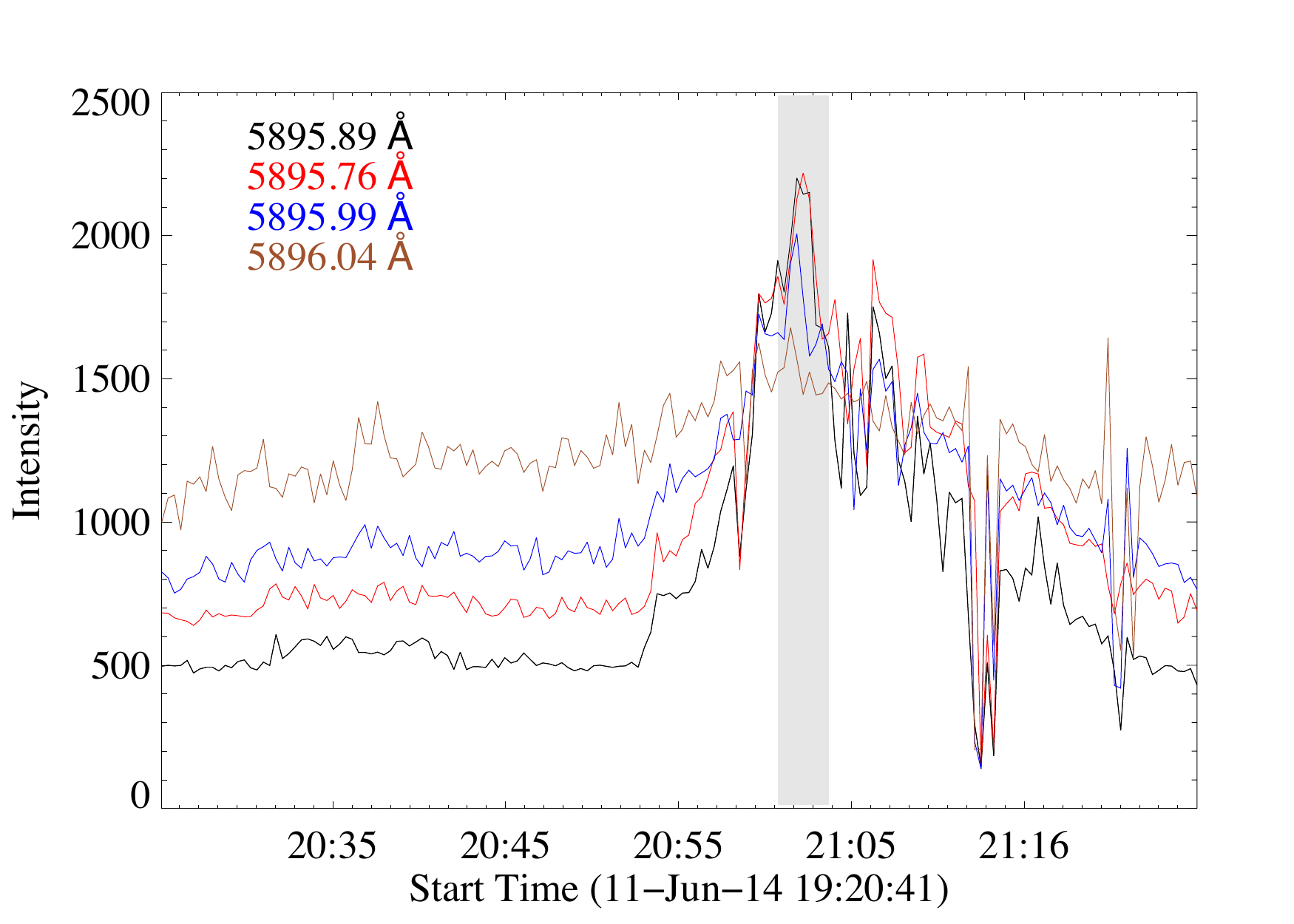}
\caption{Lightcurves of the flaring region marked with the upper blue contour in Figure 1. The grey-shaded area indicates the time interval for which line profiles presented In Figure~\ref{fig3} and \ref{fig4} 
are produced.}
\end{center}
\label{fig2}
\end{figure}

\begin{figure*}[t]
\begin{center}
\includegraphics[width=17.4 cm]{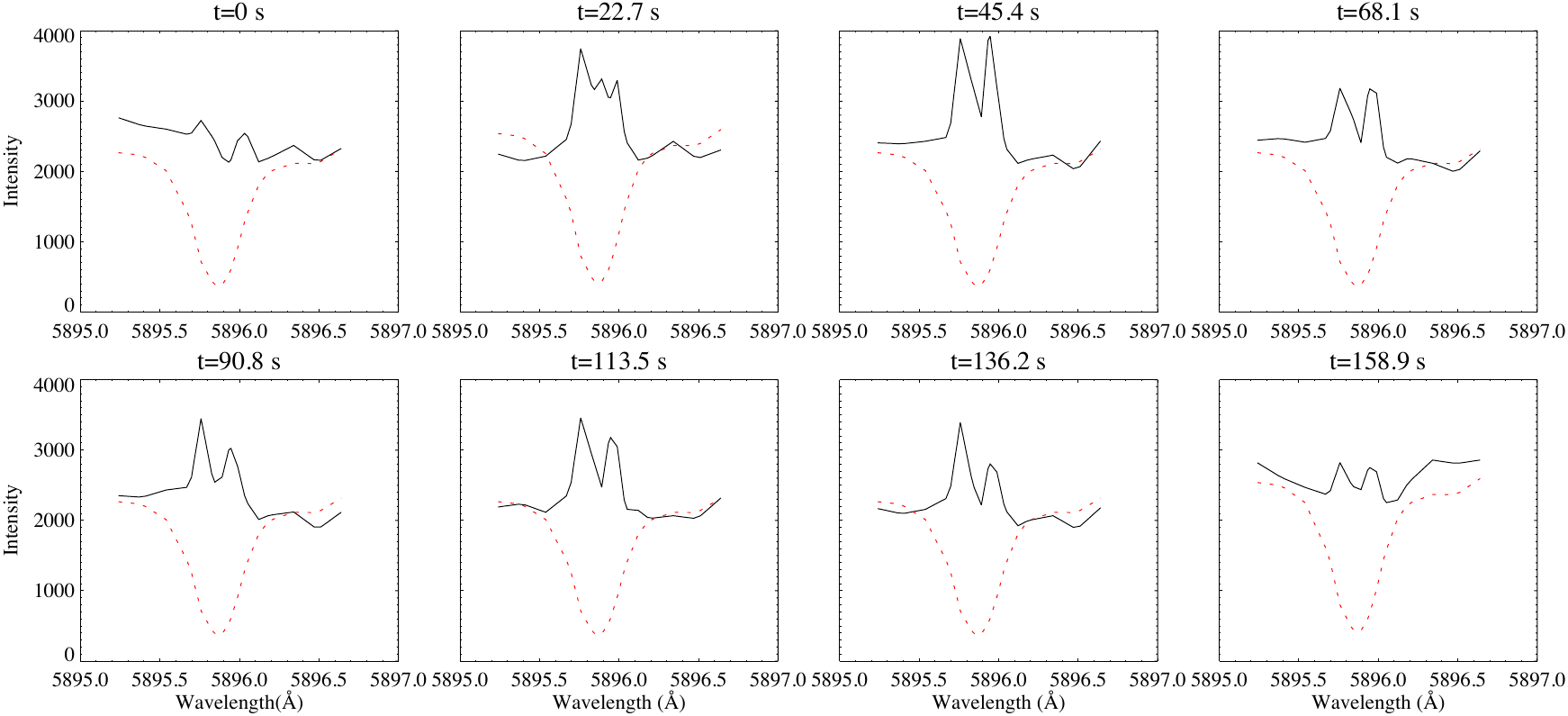}
\caption{The temporal evolution of the Na {\sc{i}} D$_1$ line profiles during flare maximum for the upper ribbon outlined with the 
blue contour in Figure~\ref{fig1}. The profiles have been obtained from the temporal interval marked with grey-shaded area in Figure~\ref{fig2}.  The red dashed lines show the mean spectrum averaged over the quiet Sun FOV for reference.}
\label{fig3}
\end{center}
\end{figure*}

The two ribbon M3.9 flare was observed in active region NOAA 12087 located 
at heliocentric coordinates $\sim(-755", -297"$).   
Figure~\ref{fig1} shows the flare images in Na {\sc{i}} D$_1$ line core and wing positions. 
Seeing was below average during the observations, but the flare-associated emission is clearly detected in the images. 
Two bright ribbons are identified in the Na {\sc{i}} D$_1$ images, highlighted with the blue contours in Figure~\ref{fig1}.
Lightcurves generated from the ribbon show the increase in the emission of the Na {\sc{i}} D$_1$ core and wings, 
peaking at 21:04 UT (Figure~\ref{fig2}).  

\begin{figure*}[t]
\begin{center}
\includegraphics[width=17.4 cm]{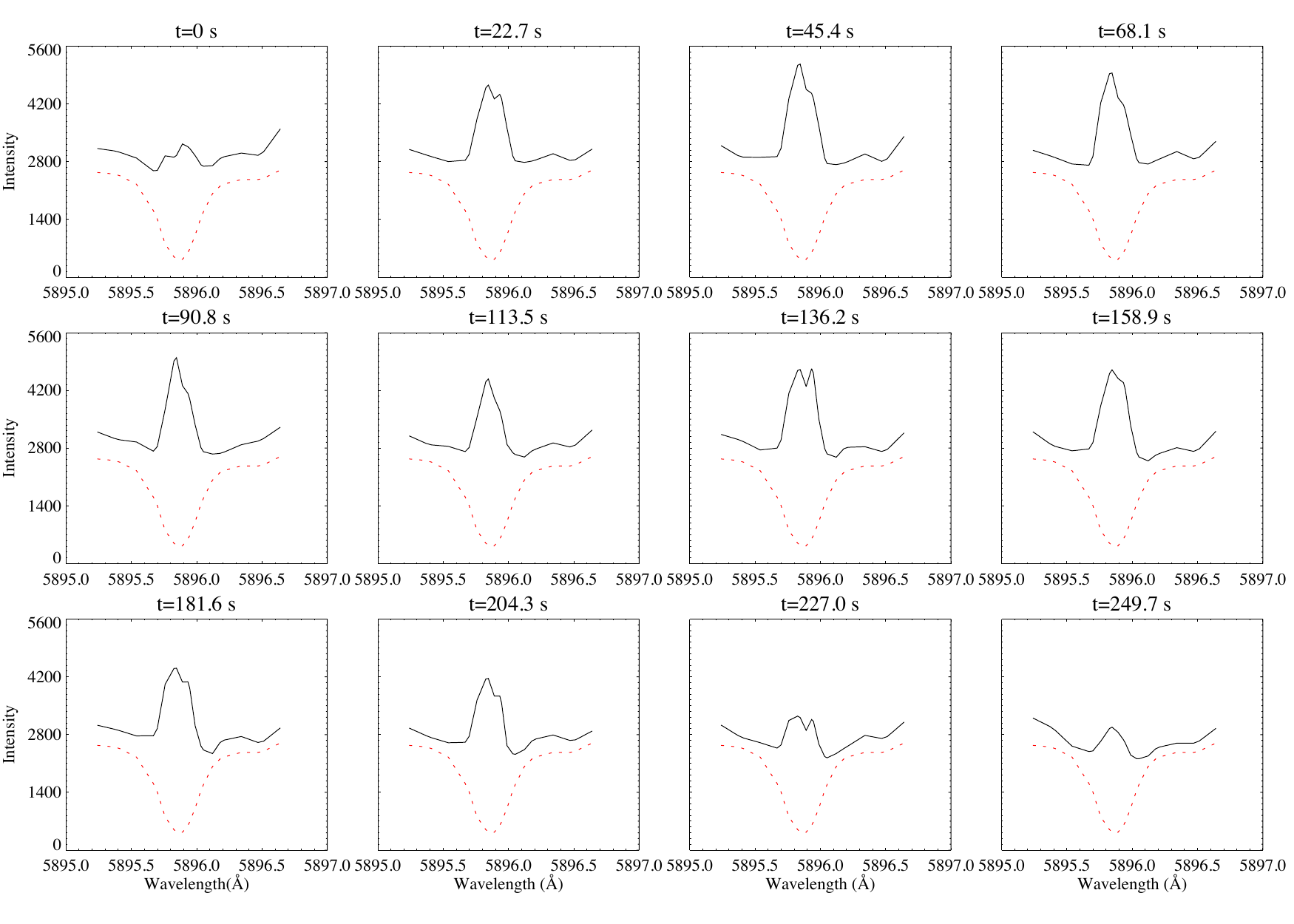}
%\center{\resizebox{20cm}{!}{\includegraphics{fig11}}}
\caption{Same as Figure~\ref{fig3}, axcept for the lower ribbon outlined with the blue contour in Figure~\ref{fig1}.} 
%The profiles have been obtained from the temporal interval marked with grey-shaded area in Figure~\ref{fig2}. 
%The red dashed lines show the mean spectrum averaged over the quiet Sun FOV for reference.}
\label{fig4}
\end{center}
\end{figure*}

In Figures~\ref{fig3} and \ref{fig4} we show the Na {\sc{i}} D$_1$ line profiles for the upper and lower ribbons respectively, 
averaged over the areas marked with the blue contour in Figure~\ref{fig1}. 
The line profiles are produced during the flare peak, which is a vertical grey shade in Figure~\ref{fig2}. 
Shortly after the flare onset, Na {\sc{i}} D$_1$ rapidly appears in emission with a central reversal. 
Furthermore, the temporal evolution of the centrally-reversed Na {\sc{i}} D$_1$ profiles shows excess emission in the blue wing 
(blue asymmetry) with nearly unshifted line center for most of the flaring profiles (Figure~\ref{fig3}, \ref{fig4}). 
After the flare maximum, from $\sim$21:07 UT, the line profile gradually changes from the centrally-reversed emission to the quiet pre-flare absorption.

%%%%%%%%%%%%%%%%%%%%%%%%%%%%%%%%%%%%%%%%%%%%%%%%%%%%%%%%%%%%%%%%%%%%%%%%%

\subsection{Hard X-rays}
\label{HXR}

\begin{figure*}
\begin{center}
\includegraphics[width=17.4 cm]{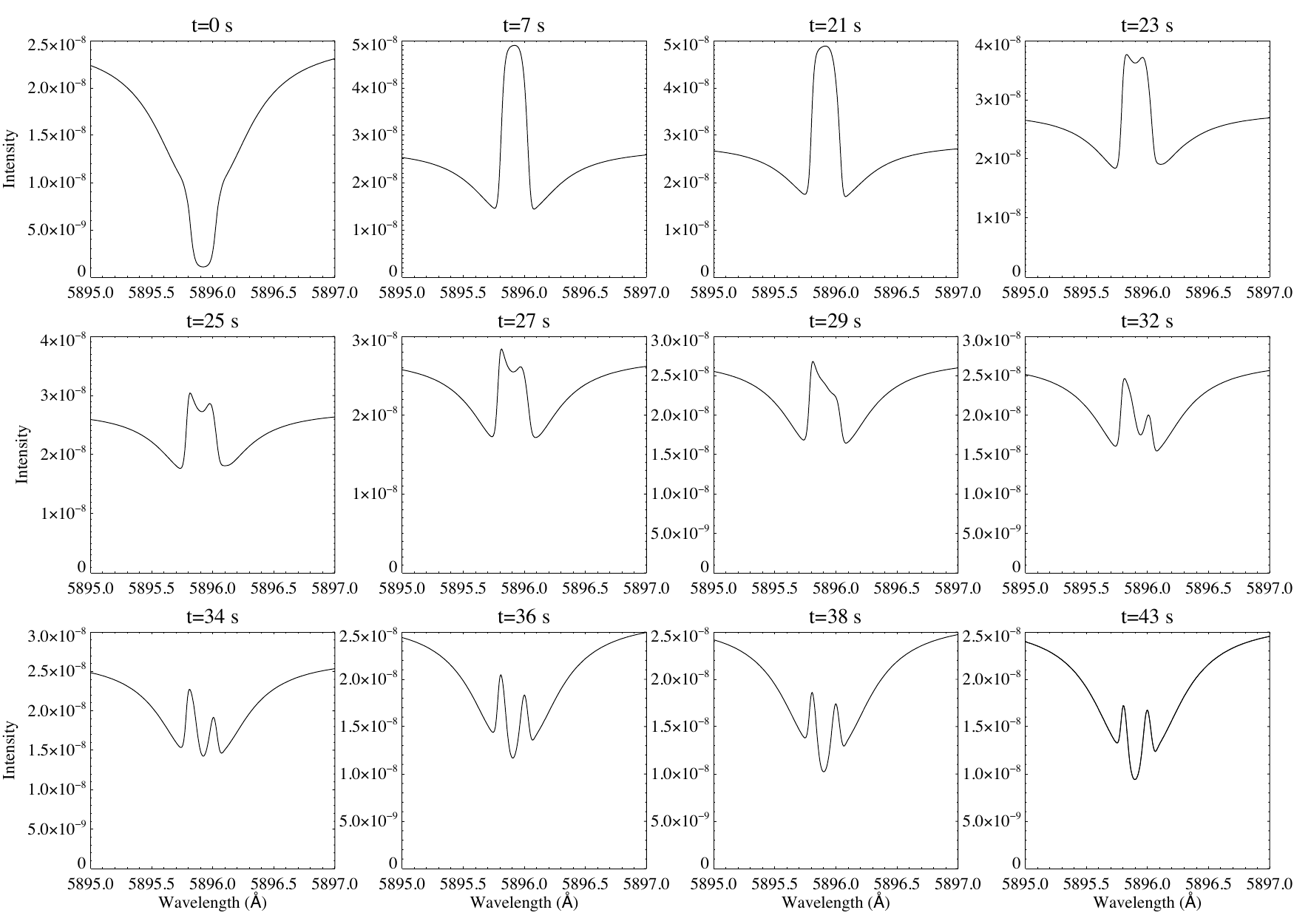}
\end{center}
\caption{The temporal evolution of the synthesized Na {\sc{i}} D$_1$ profile from the combined RADYN (F11) and RH simulations. A blue asymmetry is produced 
in the relaxation phase of flare simulation when the beam heating has seized (after t=20 s).}
\label{fig5}
\end{figure*}

The RHESSI data were fitted with an isothermal plus thick-target model, using the OSPEX function ({\tt \verb$thick2_vnorm$}), 
and calculated $P_\mathrm{nth}$ throughout the impulsive phase, integrating the counts in bins of 12 seconds. 
RHESSI detectors 5 and 7 were employed as the results from the other detectors were too noisy to be used for a reliable fit. 
The spectral results give an estimate of the energy distribution $F(E)$ of the accelerated electrons during the flare, with 
total energy $P_{th}$ derived by integrating the energy of the distribution of electrons $P_{th}= \int EF(E)dE$. %* You could cite Kuridze+2015 as an example of this method.}
We found a maximum energy rate, $P_\mathrm{nth} \approx (3.6 - 5.5)\times10^{28}$ erg s$^{-1}$, at the time of maximum of HXR emission.  
To obtain the energy flux (erg s$^{-1}$ cm$^{-2}$) deposited into the chromospheric source, the power of the non-thermal electrons 
$P_\mathrm{nth}$ must be divided by the footpoint area. The reconstructed images for the HXR emission  \citep[using CLEAN,][detectors 3-8, beam width factor of 1.5]{2002SoPh..210...61H},  
do not resolve the flare. However, the centroid of the HXR emission is well associated with the main locations of the Na {\sc{i}} D$_1$ emission.
%They only shows the HXR source as circular blob in good association with the main locations of the Na {\sc{i}} D$_1$ emission. 
Therefore, RHESSI images could not be used to estimate the size of the footpoints. 
The area of the ribbons was hence determined by contouring a region with intensity above 50\% of the maximum of the Na {\sc{i}} D$_1$ 5895.76~\AA\ wing 
image near the HXR peak, which gave a value of $\sim10^{17}$ cm$^2$. Thus, at the time of the maximum HXR emission,  
the average energy flux delivered by the electrons into the chromosphere is around $(3.6 - 5.5)\times 10^{11}$ erg s$^{-1}$ cm$^{-2}$.

%%%%%%%%%%%%%%%%%%%%%%%%%%%%%%%%%%%%%%%%%%%%%%%%%%%%%%%%%%%%%%%%%%%%%%%%%

\subsection{Simulated Na {\sc{i}} D$_1$ Line Profiles}

To interpret the observational characteristics of Na {\sc{i}} D$_1$   
we generated synthetic line profiles with the radiative-hydrodynamic code \citep[RADYN;][]{1997ApJ...481..500C, 2005ApJ...630..573A},  
and the radiative transfer code RH \citep{2001ApJ...557..389U}. RADYN snapshots generated at different time steps can be used as an input atmosphere to RH to investigate the temporal evolution of the Na {\sc{i}} D$_1$ line profile during the flare.  

\begin{figure*}[t]
%\vspace{3mm}
\begin{center}
\includegraphics[width=17.4 cm]{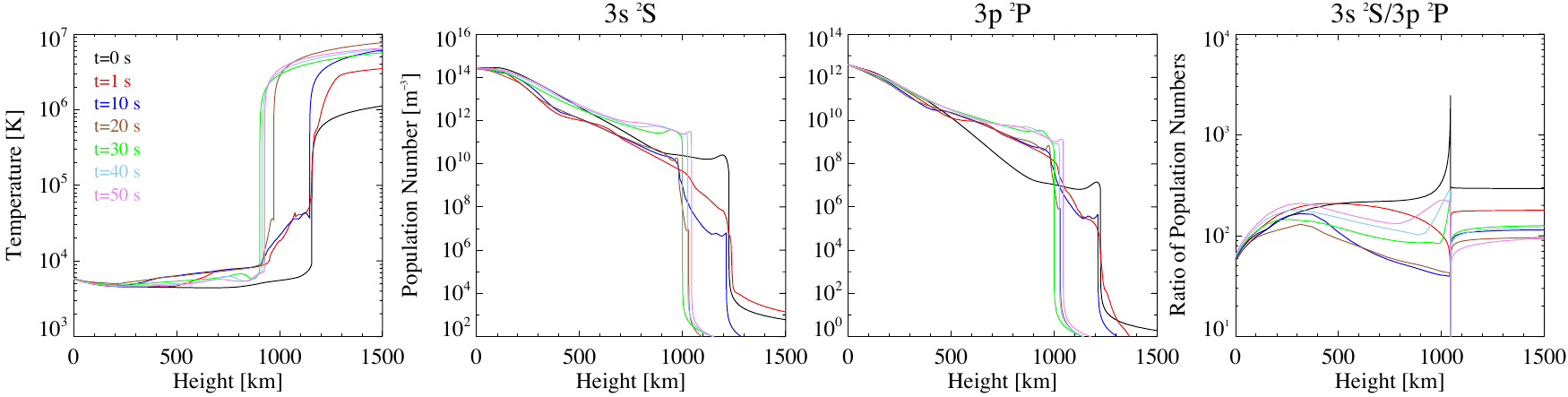}
\caption{The temperature (left panel), the density population of the energy states $3s{~^2}S$ and $3p{~^2}P$ of the sodium atom 
(middle panels) and the ratio of the densities populations of these two states (right panel) as a function of height at different stages of the simulation.}
\label{fig55}
\end{center}
\end{figure*}

The RADYN simulation was performed for a strong beam with $\mathrm{F}=10^{11}\mathrm{ergs~cm^{-2}~s^{-1}}$ (also known as an F11 flare) and an isotropic pitch angle distribution in the forward hemisphere
with the Fokker-Planck solution to the nonthermal electron distribution \citep{2015ApJ...809..104A}.
A constant heating flux was applied for 20 s, and the atmosphere was allowed to relax for an additional 40 s. 
We used a power-law index  and a low energy cut-off of  $\delta$=4.2 and $E_c$ = 25 keV, respectively.  
%The beam parameters used in the model are very close to the ones estimated from the HXR data.  
Snapshots were produced and the Na {\sc{i}} D$_1$ line profiles were synthesised 
for every single time step during the 60 s RADYN run.

Figure~\ref{fig5} shows the temporal evolution of the synthesised Na {\sc{i}} D$_1$ profiles from these combined simulations.  
Before the start of electron beam heating, at $\mathrm{t=0~s}$, the line was in absorption, but changes rapidly into emission when the beam heating is initiated. 
%This change is a result of non-thermal, impulsive phase heating.  
The evolution of the temperature profile of the flaring atmosphere shows that following the beam heating ($\mathrm{t =1~s}$), 
the temperature increases from $\sim$5000 K to $\sim\mathrm{7000-20000~K}$ at height $\sim\mathrm{500-1000~km}$ (black and red lines of the left panel of Figure~\ref{fig55}). 
The Na {\sc{i}} D$_1$ and D$_2$ lines arise due to the $\mathrm{3s^{~2}S-3p{~^2}P_{1/2,3/2}}$ transitions.
%is produced by the transition between $3s^2S\rightarrow3p^2P^{0}$ atomic energy states \citep{1992A&A...265..237B}.   
The increased temperature leads to a rapid change in the population densities of these $\mathrm{^{~2}S~and~{~^2}P}$ states % $3s^2S$ and $3p^2P^0$  
(middle and right panels of Figure~\ref{fig55}). 
In particular, the population density of the Na {\sc{i}} D$_1$ ground state ($\mathrm{3s{~^2}S}$) 
is decreased at $\mathrm{t=1~s}$ whereas the population of $\mathrm{3p{~^2}P}$ 
is increased. The right panel of Fig~\ref{fig55} shows that the ratio of the populations of the two states $\mathrm{^{~2}S/{~^2}P}$ 
decreases by a factor of 10 during the first second of beam heating of the atmosphere at a height of around $\mathrm{500-1000~km}$.  
This beam heating increase the collisional rates and excites more electrons from the ground to the first excited state. 
In turn, this increases the probability that Na {\sc{i}} D$_1$ photons, which would have previously been absorbed, escape freely and results in an increase in the line intensity.

\begin{figure*}[t]
%\vspace{3mm}
\begin{center}
\includegraphics[width=17.9 cm]{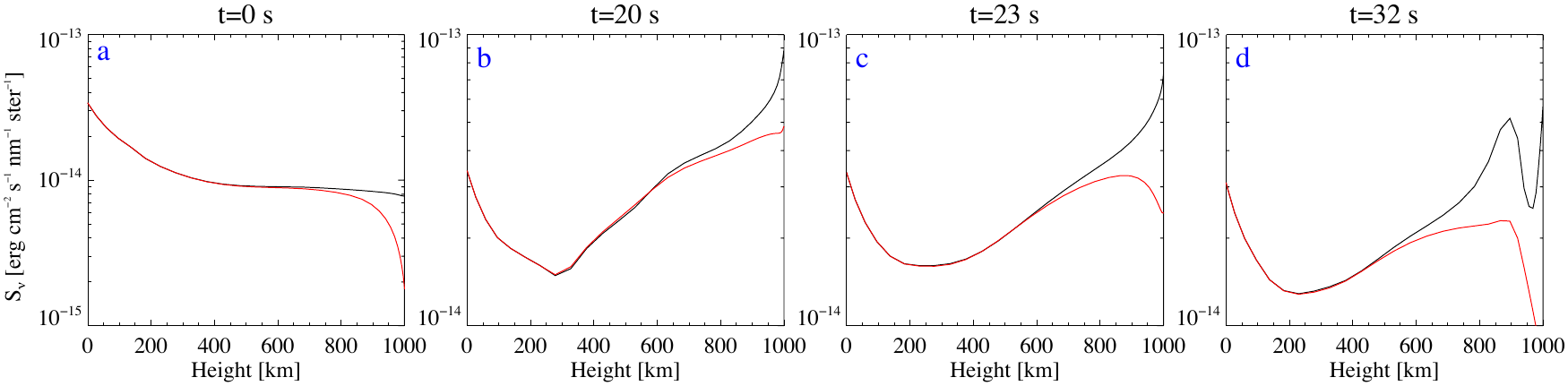}
\caption{Temporal evolution of the source function (red lines) and Planck function (black lines) at Na {\sc{i}} D$_1$ line center (5895.89 {\AA}).} 
%The temperature (left panel), the density population at the energy level $3s^2S$ and $3p^2P^{0}$ of the sodium atom (middle panels) and the ratio of the densities populations of these two levels (right panel) as a function of height at different stages of the simulation.}
\label{fig555}
\end{center}
\end{figure*}

We also investigate the line source function of the analysed RH/RADYN run to study the temporal evolution of the radiation field of the Na {\sc{i}} D$_1$ line. 
Figure~\ref{fig555} shows the source function at line center at different stage of the simulation. % of the RADYN/RH run we have analysed here.
The Planck functions (black lines) are overplotted to show the height of non-LTE decoupling in the atmosphere.
Before the start of electron beam heating, at t=0~s, the source function decreases 
with increasing height and the resulting line profile is purely in absorption (panel $a$ of Figure~\ref{fig555}). 
However, during the beam heating at a height of $\sim$300 km the source function, which in this region is strongly coupled to the Planck function, 
starts to increase production of emission from the core formation region (panel $b$ of Figure~\ref{fig555}).
The change in the source function, together with the change in the balance between the population densities of the energy states, switches the sodium line profile from absorption into emission.  
%Furthermore, an analyses of source function shows that emissions increase dramatically during the beam heating in the upper Na {\sc{i}} D$_{1}$ formation region (Figure~\ref{fig555}).}
After 20 s, when beam heating was stopped, the temperature begins to decrease and the ratio of the population densities changes to the opposite direction (Figure~\ref{fig55}). 
Furthermore, the line source function has developed a local maximum near the line core formation height at around 900 km, where the source function is decoupled from the Planck function (panels $c-d$ of Figure~\ref{fig555}). 
%This leads to increased emission in the line wings, relative to the line core, hence the centrally reversed line profiles.}
As a result, Na {\sc{i}} D$_1$ line profiles develop an absorption (central reversal) near the line core (Figure~\ref{fig5}).   
Also, after the beam heating has seized ($t = 20~s$), the centrally reversed Na {\sc{i}} D$_1$ 
line profile shows excess emission in the blue wing (blue asymmetry) with nearly unshifted line center (see panels d-h of Figure~\ref{fig55}).  This is similar to the observed line profiles.

To understand the formation of the asymmetric line profiles, we need to examine the line contribution functions \citep{1997ApJ...481..500C}. These are the intensities emitted at specific wavelengths from specific heights, 
introduced as a formal solution of the radiative transfer equation for the emergent intensity. \cite{1997ApJ...481..500C} investigated the formation of optically thick spectral line asymmetries using RADYN simulations. 
They showed that these asymmetries are produced by the velocity gradients near the formation height of the spectral line profile. 
In particular, if the velocity decreases outward (negative velocity gradient), 
then higher-lying atoms absorb at longer wavelengths (red photons), so the opacity at greater heights is smaller on 
the blue side of the line profile and a blue asymmetry is formed. Whereas, if the velocity increases outward (positive velocity gradient), 
higher lying atoms absorb at shorter wavelengths (blue photons) and the opacity is smaller on the red side of the line. 

In Figure~{\ref{fig6} we present the temporal evolution of the Na {\sc{i}} D$_1$ line contribution functions. The diagrams are plotted in red color scale, with brighter shades showing higher intensities. 
Line profiles are shown as white solid lines, while the vertical velocity structure is plotted as a red dashed line. We note that positive values for velocities correspond to plasma downflows.

\begin{figure*}[t]
\resizebox{5.03cm}{!}{\includegraphics[clip=true]{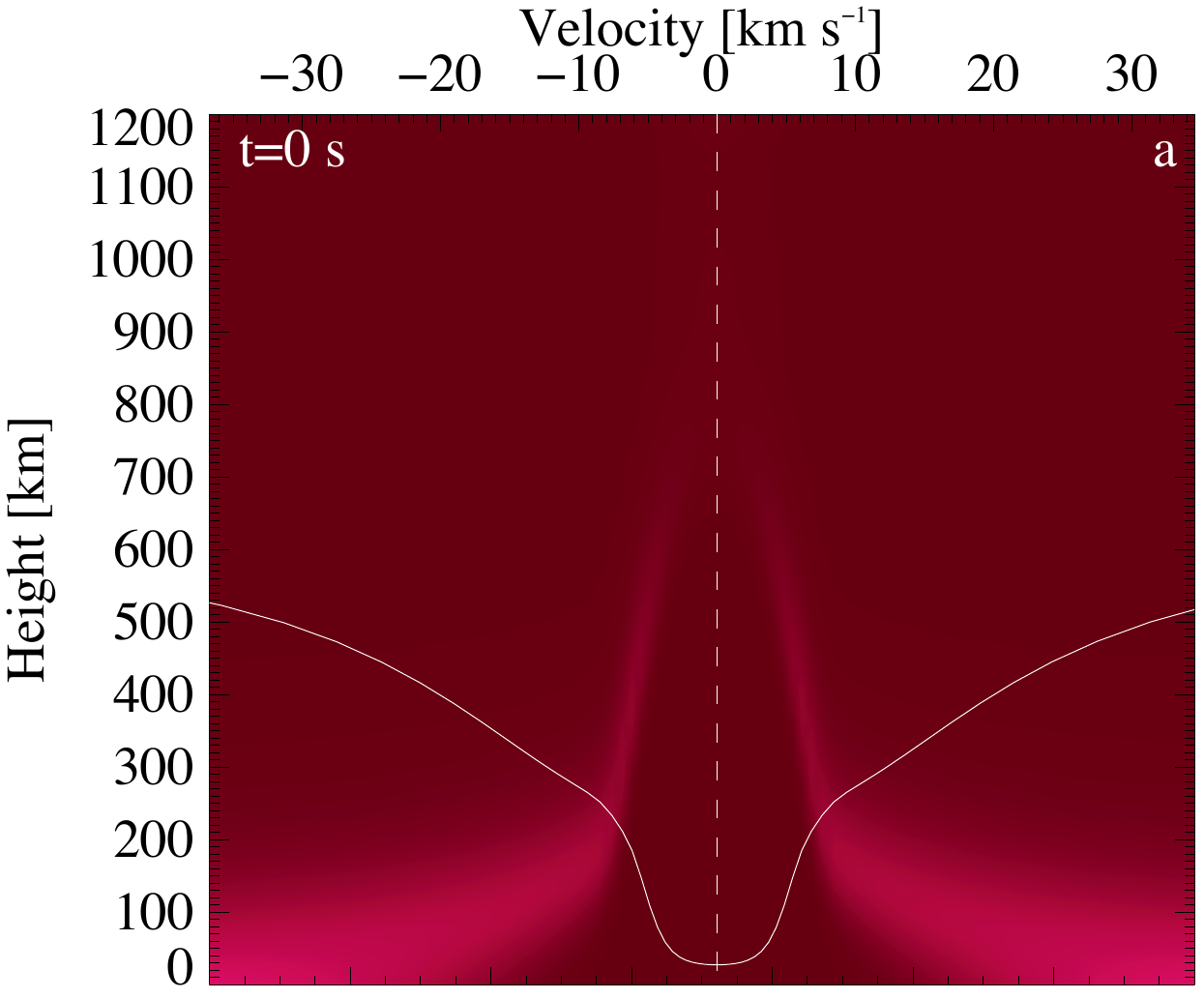}}
\resizebox{4.2cm}{!}{\includegraphics[clip=true]{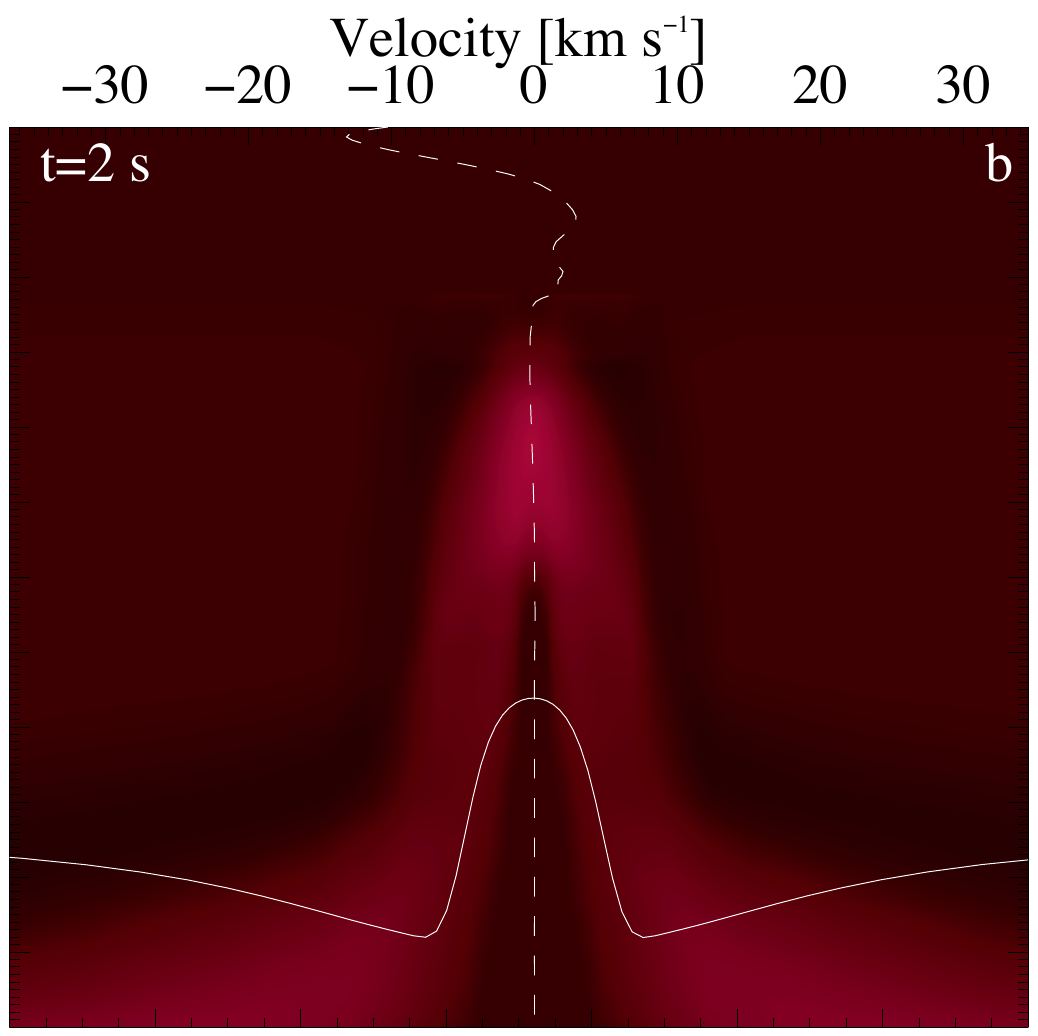}}
\resizebox{4.26cm}{!}{\includegraphics[clip=true]{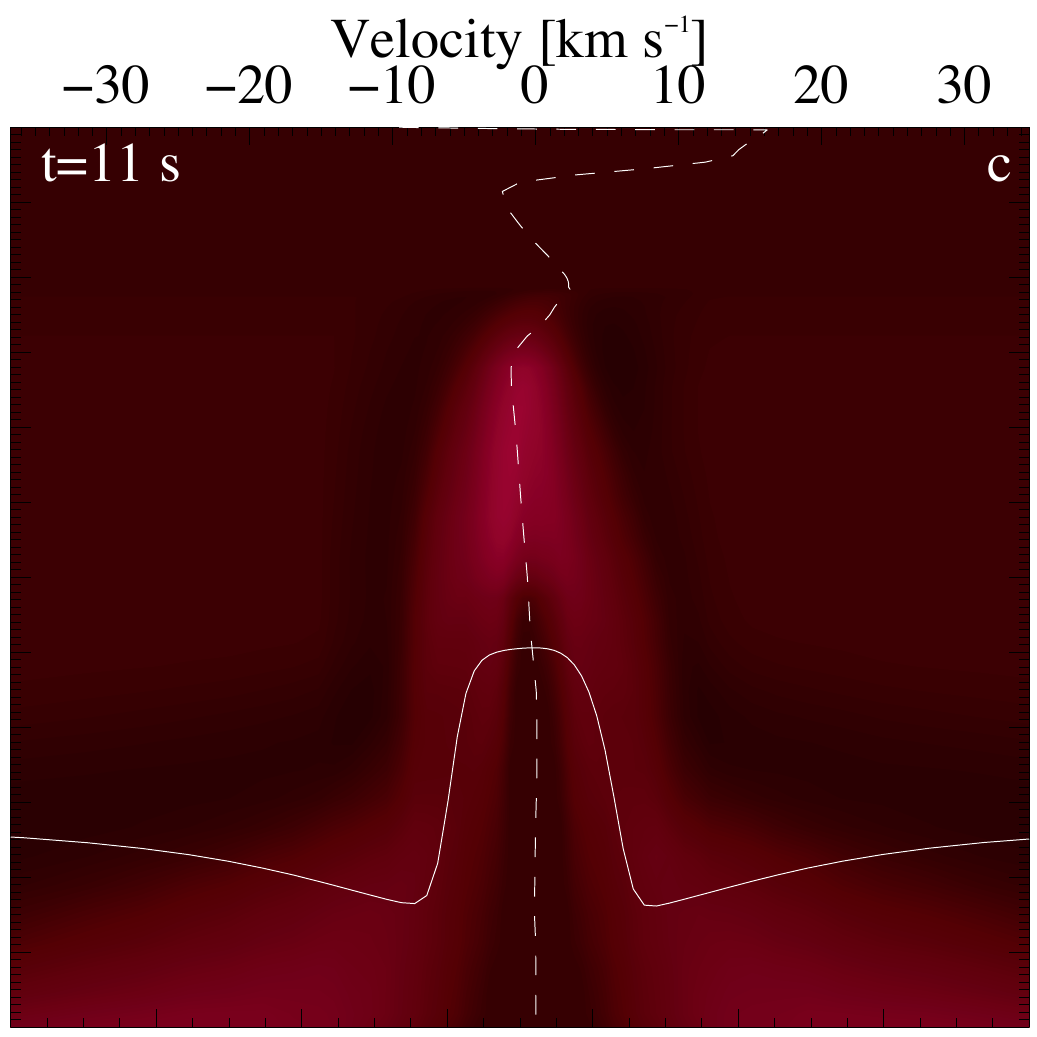}}
\resizebox{4.28cm}{!}{\includegraphics[clip=true]{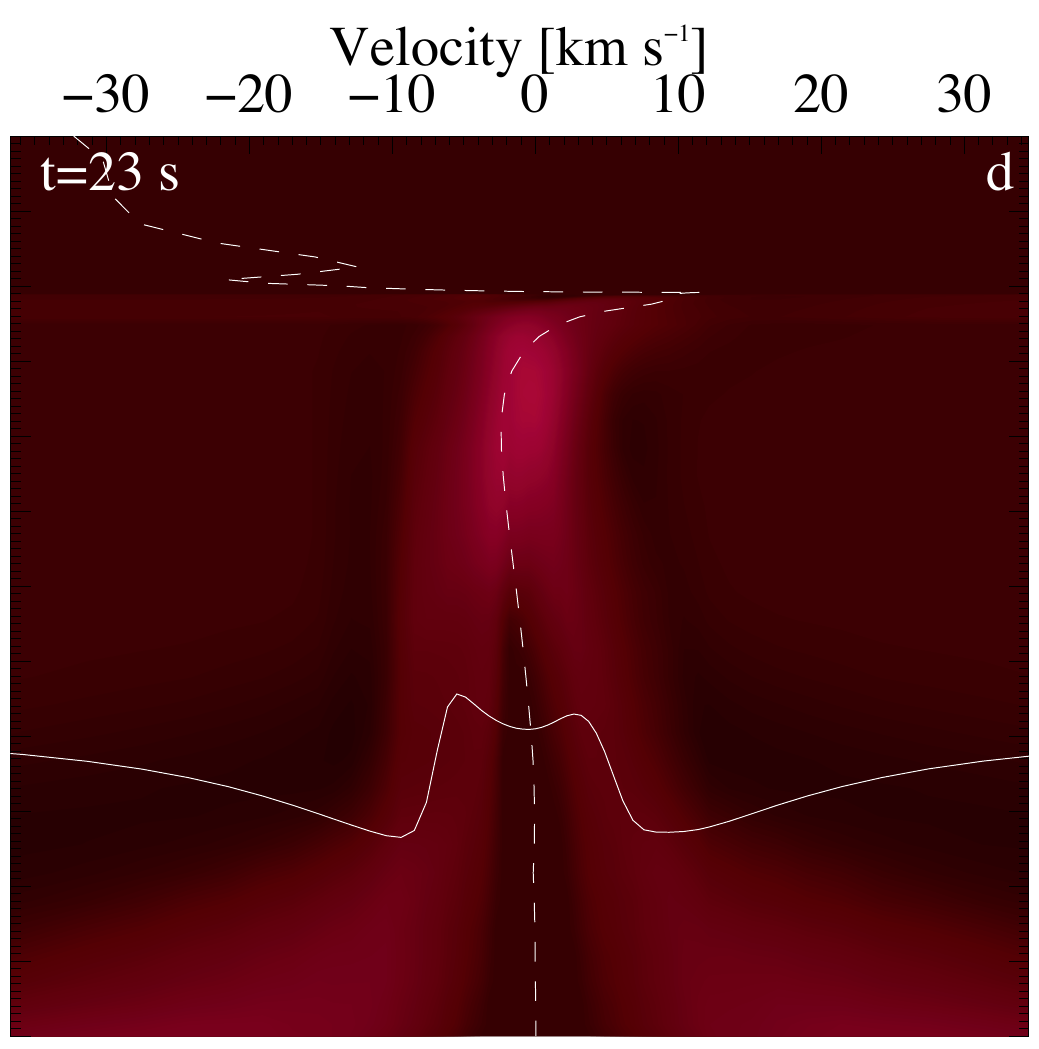}}
\resizebox{5.04cm}{!}{\includegraphics[clip=true]{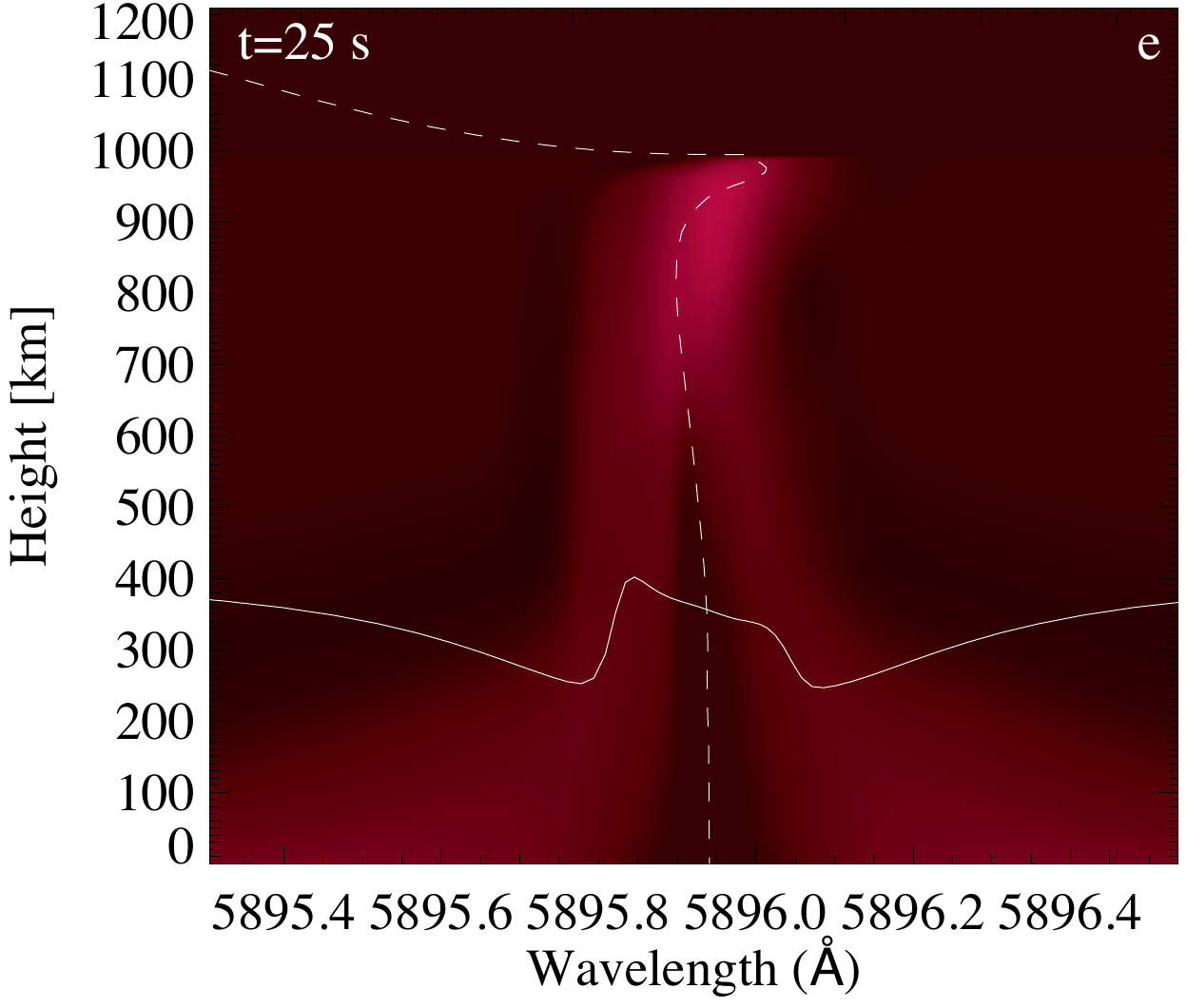}}
\resizebox{4.25cm}{!}{\includegraphics[clip=true]{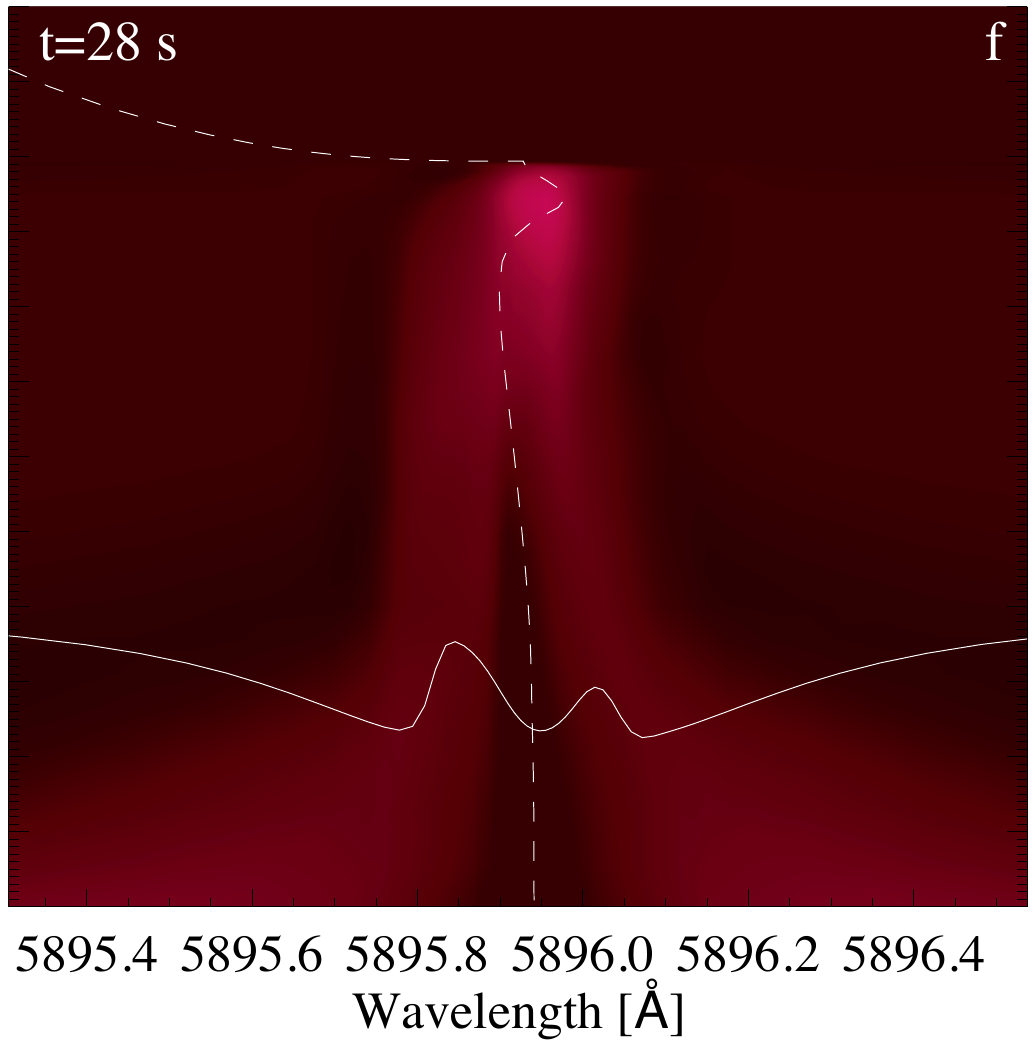}}
\resizebox{4.26cm}{!}{\includegraphics[clip=true]{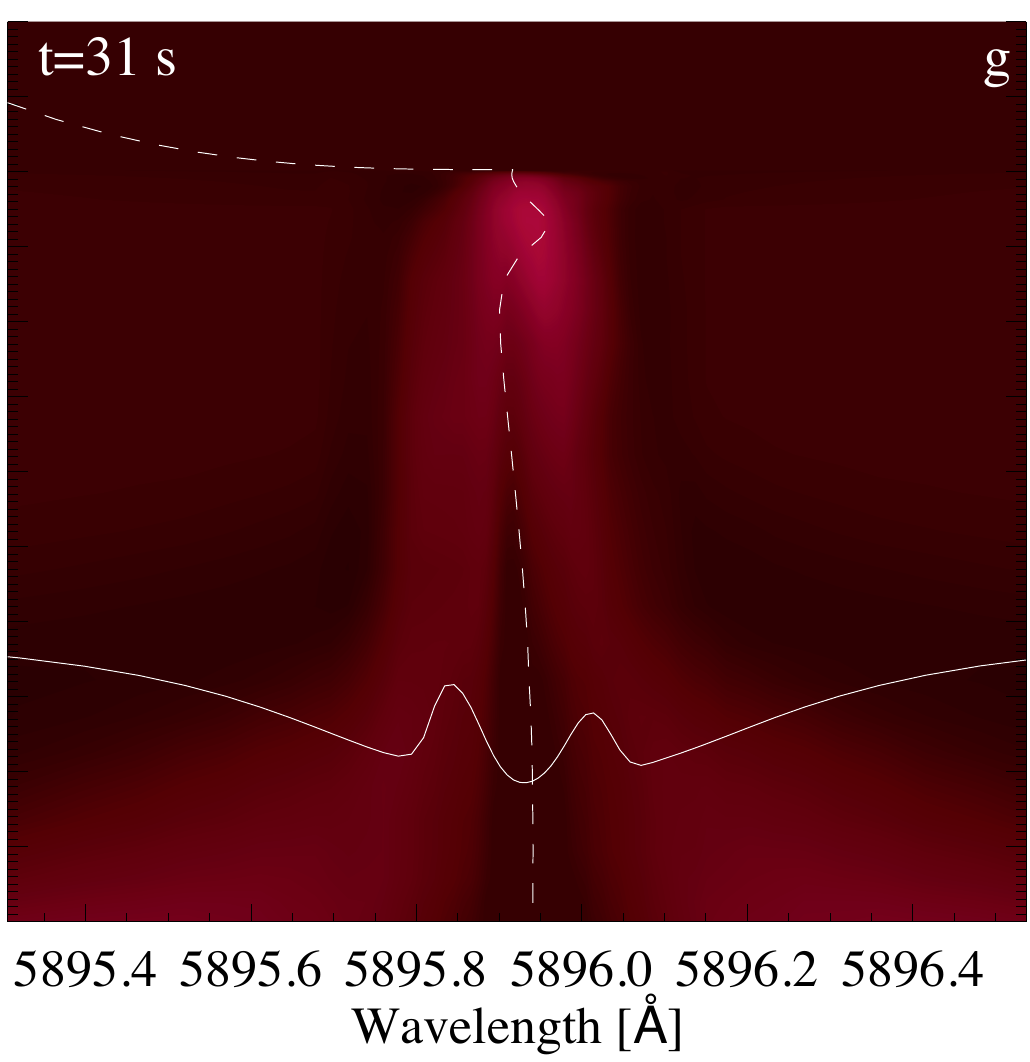}}
\resizebox{4.28cm}{!}{\includegraphics[clip=true]{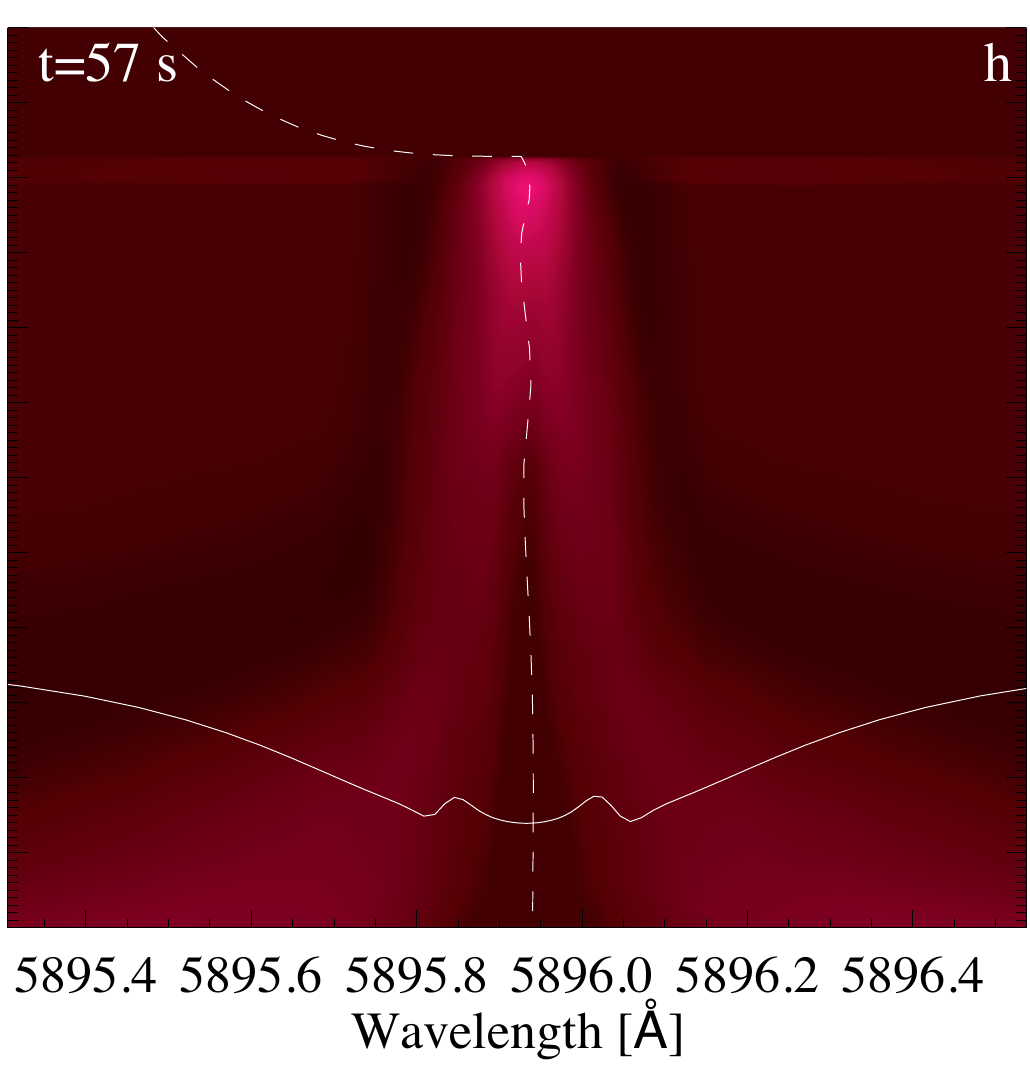}}
\caption{The intensity contribution function for the Na {\sc{i}} D$_1$ line at different stages of the simulation. 
The diagrams are plotted in a red scale so that brighter shades indicate higher intensities. The line profile is overplotted as a full white line. 
The vertical velocity structure of the plasma is overplotted as a red dashed line (negative velocity corresponds to plasma upflows).}
\label{fig6}
\end{figure*}

The contribution function diagram at $t=0$ 
shows that the Na {\sc{i}} D$_1$ line core is formed at a height of $\mathrm{\sim400 - 900~km}$ 
in the non-flaring, static atmosphere (panel $a$ of Figure~{\ref{fig6}).  
Although the electron beam disturbs the velocity field due to the non-thermal heating, during the first few seconds the 
field at the Na {\sc{i}} D$_1$ formation height is still undisturbed. This is because the primary energy release site, where strong upflows (evaporations) and downflows (condensations) are generated, 
is located above the Na {\sc{i}} D$_1$ line formation layer  % at  $<$1000~km
(Figure~{\ref{fig6}). Therefore, the Na {\sc{i}} D$_1$ line profile is still symmetric with respect to the line core.   
However, from t=20~s the lower atmosphere also has developed a weak upflow  of about  $\mathrm{2-3~\ks}$ (panel $d$ of Figure~{\ref{fig6}). 
At %$\mathrm{800-900~km}$ has positive velocity gradient whereas the lower 
$\sim\mathrm{400-800~km}$, where most of the line emission core is formed, the velocity gradient with respect to the heigh outward is negative (i. e. velocity decrease outward).
%negative velocity gradient. 
As mentioned earlier, the negative velocity gradient shifts the wavelength of maximum opacity to the red.
The higher-lying plasma of the line core absorbs red photons, increasing opacity in the red wing.
This makes the height range over which the contribution function is high greater in the blue wing of the profile, 
and the blue asymmetry is established (panel $d$ of Figure~\ref{fig6}).
The cores of the simulated Na {\sc{i}} D$_1$
profiles remain unshifted when the blue asymmetry is detected (panels $d,e,f$ of Figure~\ref{fig6}). 
This may be due to the presence of different velocity fields with positive gradients between $\sim\mathrm{850-950~km}$
which produce an effectively unshifted line core. The structure of the velocity field does not change 
during the next $\sim$20~s, and hence the blue asymmetry is maintained (panels $e,f,g$ of Figure~\ref{fig6}). 
After around 50~s, the velocity field becomes very weak and the line profile establishes a symmetric shape again (panel $h$ of Figure~\ref{fig6}).

%%%%%%%%%%%%%%%%%%%%%%%%%%%%%%%%%%%%%%%%%%%%%%%%%%%%%%%%%%%%%%%%%%%%%%%%

\subsection{H$\alpha$ vs Na {\sc{i}} D$_1$ line}

The RADYN simulations used to synthesised the Na {\sc{i}} D$_1$ line were 
also employed by \cite{2015ApJ...813..125K}
to study the evolution of the H$\alpha$ line profile of the flaring atmosphere.  
In the F11 RADYN  model, the H$\alpha$ line is formed at a height of $\mathrm{\sim900-1200~km}$ 
and hence could be used as a diagnostic of the chromosphere at heights above that of Na {\sc{i}} D$_1$ formation ($\mathrm{\sim400-900~km}$;
subsection 3.2). 
Our analysis revealed that approximately 7 -- 8 s after the beam heating has started, 
H$\alpha$ shows excess emission in the blue wing (blue asymmetry) with a redshifted line core.  
At 20 s, the beam heating maintains a temperature of $\mathrm{\sim10000-35000~K}$ (Figure~\ref{fig6}).  
This region shows a downflow velocity of $\mathrm{\sim10-15~\ks}$, with the downflow formed at t = 6 -- 7 s).  When the beam switches  off (at 20.01 s), 
the temperature in the region drops to $\mathrm{T\sim11000~K~from~T\sim35000~K~in~0.1~s}$.  The pressure ratio between the material that 
suddenly cooled $(P_{below~TR})$ and the maximum pressure in the flare transition region $(P_{TR})$ changes rapidly: $P_{TR} / P_{below~TR}$ increases.  
This drives another downflow at a higher velocity of $\mathrm{v\sim20-25~\ks}$.  The two downflows are cooling rapidly as this 
material increases in density, while the beam heating is no longer present to slow the cooling. Material just below 
the flare transition region decreases in velocity, while the temperature decreases to $\mathrm{6500-7000~K~at~t=27~s}$. 
Thus, the downflowing material is sufficiently hot to excite the n = 2 level of hydrogen when the beam heating is on $\mathrm{(t<20~s)}$. 
%{\underline DATO THE LAST SENTENCE IS NOT CLEAR. YOU HAVE TRANSITIONS FROM n=2 TO WHAT LEVELS? DO YOU MEAN HIGHER LEVELS THUS CREATING THE ABSORPTION?}

On the other hand, for $\mathrm{t<20~s}$ the temperature of the material at lower heights ($\mathrm{\sim400-850~km}$), which does not have appreciable velocity (small upflows),  
is $\sim{7000-7700~K}$ and emits in Na {\sc{i}} D$_1$ (Figure~\ref{fig6}). However, when the beam heating turns off 
the temperature in this region drops, with the value of the drop increasing 
with height in this range.  
Therefore the pressure ratio between the material in the upper and lower areas in the $\mathrm{\sim400-850~km}$ region decrease.     
This drives the upflow with a velocity of $\mathrm{v\sim2-3~s~\ks}$ which in turn produces the blue asymmetry in the Na {\sc{i}} D$_1$ line profie.

%%%%%%%%%%%%%%%%%%%%%%%%%%%%%%%%%%%%%%%%%%%%%%%%%%%%%%%%%%%%%%%%%%%%%%%%%

\section{Discussion and conclusion}

We have presented spectroscopic observations of the Na {\sc{i}} D$_1$ 
line in an M3.9 flare and compared our findings with radiative hydrodynamic simulations. 
Our high spectral resolution observations show that during the flare 
the Na {\sc{i}} D$_1$ line goes into emission and a central reversal is formed (Figures~\ref{fig3} and 4). 
The analysis of synthetic line profiles indicates that the change from absorption to emission 
is a result of the heating of the lower solar atmosphere by the non-thermal electron beam. 
Although the formation of the line occurs deep below the primary non-thermal energy dissipation site,   
it is still strongly affected by the heating as the temperature of the region responds immediately to the electron beam (left panel of Figure~\ref{fig55}).  
The heating rapidly changes the balance between the population densities of the energy states in the sodium atom (Figure~\ref{fig55}), 
with increased collision rates exciting electrons from the ground $\mathrm{^{~2}S}$  
to the first excited state $\mathrm{{~^2}P}$ which allows Na {\sc{i}} D$_1$ photons to escape freely and results in an increase in the line intensity.
Furthermore, the line source function shows a local minimum near 300 km and increases toward the core formation height (Figure~\ref{fig555}).
As a result, during the heating phase the line profile developes fully into emission.
However, when the beam heating was stopped the temperature starts to drop and the ratio of the population densities changes back again (Figure~\ref{fig55}). 
The line source function has developed a local maximum near 900 km, in the region where the source function is already 
decoupled with the Planck function.
%However, later after the beam heating has seized, 
As a result, the line profiles develop a small absorption near line core (central reversal), as shown in Figure~\ref{fig5}.  
The observed line profiles have similar centrally-reversed shapes during the flare (Figures~\ref{fig3} and \ref{fig4}), 
with the absorption dip smaller for the lower ribbon profiles, indicating that the heating was more intense in the lower ribbon (Figures~\ref{fig3} and \ref{fig4}). 
%Indeed, an average intensity is bigger in the lower ribbon. 
Indeed, RH simulation confirms that in the flaring atmosphere produced with the lower electron flux (e.g. F9 model),
the Na {\sc{i}} D$_1$ profiles has a deeper absorption dip at the relaxation phase.

It must be noted that the central reversal in the synthetic Na {\sc{i}} D$_1$ line profile 
appears only after heating ends, i.e., in the relaxation phase of the flare simulation.
During the beam heating the line profile develops fully into emission without reversal.  
Recent high spatial resolution ground-based observations of solar flares in H$\alpha$ and He {\sc{i}} 10830 {\AA} lines 
indicating that the ribbon kernels of the M-class solar flares could have a very narrow width, $<$500~km \citep{2016ApJ...819...89X, 2016NatSR...624319J}.  %(Xu et al., 2016, Ap.J., 819, 89, Jing et al., 2016, Nature Scientific Reports, 6, 24319), is.
This suggest that the actual footpoint size of the flaring loops could be smaller, and hence, energy flux could be higher than an estimated 10$^{11}$ erg s$^{-1}$ cm$^{-2}$.
To assess the effects of a higher energy input on the synthesised Na {\sc{i}} D$_1$ line profile
we performed RADYN simulations for a stronger ($\mathrm{5\times~F11}$) energy flux and used resulted atmospheric snapshots in RH for synthesising Na {\sc{i}} D$_1$ line profiles.   
The obtained line profiles shows a similar evolution patterns, in particular,  
when beam heating is on Na {\sc{i}} D$_1$ is in total emission (without central reversal) and when beam heating is off the line profiles are centrally-reversed with asymmetric wing emission depending of the velocity field. 
Similar behaviour has been shown by the synthesised Na {\sc{i}} D$_1$ spectra simulated with RH for a weaker (F9) RADYN atmosphere.
Therefore, the response of the atmosphere at the Na {\sc{i}} D$_1$ formation height 
to the different energy beam heating is qualitatively similar (however, stronger ($\mathrm{F12}$) case have to be examined as well.)
This shows that the formation of the central reversal in the spectra 
could be used as a diagnostic of the non-thermal heating processes in solar flares.

%Furthermore, blue peak-to-peak asymmetry is in relaxation phase of flare simulation without beam heating
The temporal evolution of the line profile shows excess emission in the blue wing with an almost unshifted line core  (Figures~\ref{fig3} and 4). 
%The asymmetries in optically thick spectral lines of the flaring atmosphere have been known for a long time and usually attributed to the downflow and upflows of plasmas.
In an atmosphere without velocity fields, the centrally-reversed chromospheric line profiles are symmetric with respect to the line core \citep{1993A&A...274..917F,2006ApJ...653..733C}. However,  
dynamic models account for the mass motions of the flaring material and reproduce the asymmetric signatures seen in the observations.
Our Figure~\ref{fig6} shows that in a zero velocity field  the line profile is indeed symmetric. The velocity field in the lower solar atmosphere is not disturbed during the first 
10 s of the active beam heating that produces the explosive evaporation above 1000 km, which is higher than line formation height. 
However, after 20 s the lower atmosphere has developed upflows of around $\mathrm{2-3~\ks}$,
which are generally attributed to gentle evaporation. As a result, the symmetry in the line profiles is now broken. 
The negative velocity gradient at around $\mathrm{400-800~km}$ 
modifies the optical depth of the atmosphere in such a way that higher-lying (core) atoms absorb photons 
with longer wavelengths (red wing photons) and the blue asymmetry is formed (Figure~\ref{fig6}).

The line cores of the observed and simulated Na {\sc{i}} D$_1$ asymmetric line
profiles remain unshifted (panels $d,e,f$ of Figure~\ref{fig6}), in contrast to the centrally-reversed 
H$\alpha$ line profiles which show a red-shifted line core during the blue asymmetry \citep{2015ApJ...813..125K}. 
This may be due to more complex velocity fields with different condensation/evaporation patterns.
Indeed, Figure~\ref{fig6} shows that the velocity gradient changes sign above 800 km, which covers the upper, narrow layer of core formation height. This can produce an effectively unshifted line core. 

To our knowledge, we have presented the first high spectral resolution imaging spectroscopy of a solar flare in the Na {\sc{i}} D$_1$ line. 
The simulated line profiles show good agreement with observations, indicating that they can be a very important diagnostic 
of the properties and dynamics of the lower flaring atmosphere 
located below the formation height of the H$\alpha$ and Ca {\sc{ii}} line cores. %, simultanios spectral observations in these lines can provide full .  }
%Therefore, further observations with higher temporal, spatial and spectral resolution observations and more realistic simulations, is recommended.
We have shown that as in H$\alpha$, the asymmetries in centrally-reversed Na {\sc{i}} D$_1$ spectral profiles could be 
an effective tracer of the velocity field in the flaring atmosphere.

%%%%%%%%%%%%%%%%%%%%%%%%%%%%%%%%%%%%%%%%%%%%%%%%%%%%%%%%%%%%%%%%%%%%%%%%%

\begin{acknowledgements}
The Dunn Solar Telescope at Sacramento Peak/NM is operated by the National Solar Observatory (NSO). NSO is operated by the Association of Universities for Research in Astronomy (AURA), 
Inc. under cooperative agreement with the National Science Foundation (NSF). The research leading to these results has received funding from the European Community's 
Seventh Framework Programme (FP7/2007-2013) under grant agreement no. 606862 (F-CHROMA).
\end{acknowledgements}

\bibliography{bibtex.bib}
\end{document}